\documentstyle[12pt]{article}

\def\overlay#1#2{\ifmmode \setbox 0=\hbox {$#1$}\setbox 1=\hbox to\wd 0{\hss
$#2$\hss }\else \setbox 0=\hbox {#1}\setbox 1=\hbox to\wd 0{\hss #2\hss }\fi
#1\hskip -\wd 0\box 1}

\catcode`@=11
\newcount\@tempcntc
\def\@citex[#1]#2{\if@filesw\immediate\write\@auxout{\string\citation{#2}}\fi
  \@tempcnta\z@\@tempcntb\m@ne\def\@citea{}\@cite{\@for\@citeb:=#2\do
    {\@ifundefined
       {b@\@citeb}{\@citeo\@tempcntb\m@ne\@citea\def\@citea{,}{\bf ?}\@warning
       {Citation `\@citeb' on page \thepage \space undefined}}%
    {\setbox\z@\hbox{\global\@tempcntc0\csname b@\@citeb\endcsname\relax}%
     \ifnum\@tempcntc=\z@ \@citeo\@tempcntb\m@ne
       \@citea\def\@citea{,}\hbox{\csname b@\@citeb\endcsname}%
     \else
      \advance\@tempcntb\@ne
      \ifnum\@tempcntb=\@tempcntc
      \else\advance\@tempcntb\m@ne\@citeo
      \@tempcnta\@tempcntc\@tempcntb\@tempcntc\fi\fi}}\@citeo}{#1}}
\def\@citeo{\ifnum\@tempcnta>\@tempcntb\else\@citea\def\@citea{,}%
  \ifnum\@tempcnta=\@tempcntb\the\@tempcnta\else
   {\advance\@tempcnta\@ne\ifnum\@tempcnta=\@tempcntb \else \def\@citea{--}\fi
    \advance\@tempcnta\m@ne\the\@tempcnta\@citea\the\@tempcntb}\fi\fi}
\catcode`@=12

\begin{document}

\font\fortssbx=cmssbx10 scaled \magstep2
\hbox to \hsize{
\hskip.5in \raise.1in\hbox{\fortssbx University of Wisconsin - Madison}
\hfill\vtop{\hbox{\bf MAD/PH/790}
                \hbox{\bf RAL-93-093}
                \hbox{\bf TIFR/TH/93-39}
                \hbox{November 1993}}}

\vspace{.5in}

\begin{center}
{\large\bf HEAVY CHARGED HIGGS SIGNALS\\ AT THE LHC}\\[.2in]
V.~Barger$^a$, R.J.N.~Phillips$^b$, and D.P.~Roy$^c$\\[.1in]
{\it
$^a$Physics Department, University of Wisconsin, Madison, WI 53706, USA\\
$^b$Rutherford Appleton Laboratory, Chilton, Didcot, Oxon OX11 0QX, UK\\
$^c$Tata Institute of Fundamental Research, Bombay 400005, India}
\end{center}

\vspace{.5in}
\begin{abstract}

We discuss the viability of $gb \to tH \to ttb$ charged-Higgs signals
at the proposed LHC $pp$ supercollider, in the decay
channel  $tt \to (bq \bar q')(b \ell \nu)$.  Here one top quark decays
hadronically and one semileptonically, with all three $b$-quarks giving
flavor-tagged jets.  The principal backgrounds come from $ttg,ttq,ttc$
and $ttb$ continuum production, with possible mis-tagging of $g,q$ and
$c$.  We conclude that significant signals can be separated from these
backgrounds, for limited but interesting ranges of the parameters
$m_{H\pm}$ and $\tan\beta$, with the LHC energy and luminosity.
\end{abstract}

\thispagestyle{empty}

\newpage

   The search for Higgs bosons is in the forefront of present research
effort in particle physics\cite{hhg}.  While there is a single Higgs
boson in the Standard Model (SM), the minimal supersymmetric extension
(MSSM) has five of them --- three neutral $(h,H,A)$ and two charged
$(H^{\pm})$.  Phenomenological interest here has concentrated largely
on the neutral sector\cite{baer,gunion,kz,barger}. As regards $H^{\pm}$,
it is recognized that top decay would provide viable signals at
hadron colliders if $m_{H^\pm} < m_t$\cite{kz,barger,bhp,bawa,%
roy,barnett}.  On the other hand, the region $m_{H^\pm} > m_t$ is favored by
constraints from  $b \to s\gamma$  data\cite{bsg}, if there are no light
charginos\cite{bertolini}; this region has been
considered problematical, since the principal signal $H \to tb$ would
suffer from large QCD backgrounds at a hadron collider\cite{bawa,gkw}.
However, the possibility of efficient $b$-tagging could transform this
situation by discriminating against the background, as in the case of
neutral Higgs signals in the intermediate mass region\cite{gkw,dduw,dai}.
The present letter is devoted to a quantitative exploration of this
possibility; our results apply to two-Higgs-doublet models
in general, though we shall refer to particular features of the MSSM from
time to time.  Some preliminary results from a similar study by
Gunion\cite{gun2} have recently appeared; these are complementary to
the present work, since his methods of calculation and analysis differ
somewhat from ours.    We show below that viable signals may indeed
be expected, over a limited but interesting range of $H^{\pm}$ mass and
coupling parameter space, in the proposed Large Hadron Collider (LHC)
\cite{lhc} with $pp$ collisions at CM energy $\sqrt s = 14$~TeV.

   In two-Higgs-doublet models, where it is usually
assumed that up-type and down-type quarks get masses
from different vevs, the main $H^{\pm}$ interactions with quarks are given by
\begin{equation}
L = \frac{gV_{tb}}{2\sqrt 2 M_W}
      H^+ t \Bigl[ m_t \cot\beta \left(1-\gamma_5\right) + m_b \tan\beta
\left(1 + \gamma_5\right) \Bigr] b + \rm h.c. \;,
\end{equation}
neglecting terms suppressed by small quark masses or small KM matrix
elements $V_{ij}$, where $\tan\beta = v_2/v_1$ is the usual ratio of vevs. The
principal hadroproduction and decay mechanisms for a heavy charged Higgs
boson are therefore
\begin{equation}
    gb \to tH^- \to t \bar t b  \to W^+ W^- b b \bar b \;,
\end{equation}
plus the corresponding charge-conjugate channel.
(In the MSSM, an alternative decay mode
to the same final state,  $H^- \to W^-h \to W^- b \bar b$, is suppressed in
the mass range $m_{H^\pm} > m_t$ of present interest\cite{hhg}).  As a tag
for top production, we shall assume that one of the $W$-bosons decays
leptonically $W \to \ell\nu$ (with $\ell = e,\mu$).  To enhance the event
rate and facilitate event reconstruction, we assume that the other
$W$-boson decays hadronically  $W \to q\bar q'$, with invariant mass
$m(q\bar q') \simeq M_W$.  Thus we consider the signal
\begin{equation}
   gb \to tH \to bbbqq'\ell\nu \;,
\end{equation}
where all five quarks give separate jets and the lepton is isolated. We also
assume that all three
$b$-jets are tagged by a vertex detector;  tagging via semileptonic $b$-decays
is less desirable, since the additional missing neutrinos blur
the kinematics, but on the other hand it distinguishes $b$ from $\bar b$ and
removes some ambiguity in the event reconstruction. This final state implies a
spectator $b$-quark in one of the beams; however, we expect that this spectator
will be produced at small angle and will not appear in the acceptance
region described below.  Our approach differs here from Gunion\cite{gun2}
who calculates the subprocess $gg \to tbH$ where the spectator is
explicit.

   The principal background sub-processes are QCD production
\begin{equation}
     gb \to t\bar tb     \label{QCDprod}
\end{equation}
and fake backgrounds from
\begin{equation}
    gg, q\bar q \to t\bar tg \;, \qquad gq \to t\bar tq \;,  \label{fake}
\end{equation}
where the $g(q)$ jet or one of the $W \to qq'$ jets is mistakenly tagged;
$tt \to bbWW \to bbqq'\ell\nu$ decays are understood.  There is an
electroweak contribution to Eq.(\ref{QCDprod}) from $H^{\pm}$ exchange in the
$t$-channel, but this is much smaller than the signal (suppressed by
additional propagators) and we henceforth neglect it.
There is also a possible background from
intermediate-mass neutral Higgs boson production and decay:
\begin{equation}
 gg \to t\bar t H^0 \to t\bar t b \bar b \;,  \label{i-m H^0}
\end{equation}
where one of the final $b$-quarks does not give a
separate jet within acceptance cuts.  In the MSSM,
this neutral boson could be $h$ or $H$ or $A$; with
our present heavy $H^{\pm}$ scenario, we would then
have $H$ and $A$ equally heavy $(m_{H^\pm} \sim m_H \sim m_A$
with their $bb$ contributions suppressed by competing
channels $H \to hh,\ WW$ and $A \to Zh$) while $h$ couplings
are approximately those of the SM.
However, the total $tth$ production \cite{barger} is then an order
of magnitude smaller than $ttb$ production via Eq.(\ref{QCDprod}),
so we henceforth neglect the channel of Eq.(\ref{i-m H^0}).

It is already known\cite{bawa,gkw} that these backgrounds are
potentially much larger than the signal.  However, we shall show that
the background of Eq.(\ref{QCDprod}) can be reduced to the same order as
the signal  (in favorable cases) by a choice of kinematic cuts, while
the fake background Eq.(\ref{fake}) is also reduced to
a comparable level by the additional $b$-tagging requirement.
We here choose the following acceptance cuts on the 3 tagged plus
2 untagged jets (collectively labelled $j$), the lepton $\ell$ and
missing transverse momentum $\overlay/p_T$:
\begin{eqnarray}
   p_T(j), p_T(\ell), \overlay/p_T  &>& 30~\rm GeV \;, \\
   |\eta(j)|, |\eta(\ell)| &<& 2.0 \;,
\end{eqnarray}
where $p_T$ and $\eta$ denote transverse momentum and pseudorapidity.  We
also require minimum separations $\Delta R = \left[(\Delta \phi)^2 +
(\Delta \eta)^2\right]^{1/2}$ between the jets and lepton,
\begin{equation}
   \Delta R(jj) \;, \Delta R(j\ell) > 0.4 \;,
\end{equation}
to simulate some effects of jet-finding  and lepton isolation criteria.
We take account of possible invisible neutrino
energy in $b \to c \to s$ decays by Monte Carlo
modelling, and thereafter regard all partons
as jets if they pass the above cuts.
We simulate calorimeter resolution by a gaussian smearing of $p_T$,
with $(\sigma(p_T)/p_T)^2  = (0.6/\sqrt{p_T})^2 + (0.04)^2$ for jets and
$(\sigma (p_T)/p_T)^2  = (0.12/\sqrt{p_T})^2 + (0.01)^2$
for leptons (taking the same resolution for $e$ and $\mu$ for simplicity).
The $\overlay/p_T$ is evaluated from the vector sum of lepton and jet
momenta, after resolution smearing. We require the invariant mass of the two
untagged  jets to be consistent with $M_W$:
\begin{equation}
                     | m(qq') - M_W | < 15\rm~GeV
\end{equation}
We assume branching fractions $B(t \to bqq') = 2/3,\ B(t \to b\ell\nu)=2/9$,
and tagging efficiencies $\epsilon_b = 0.30,\ \epsilon_c = 0.05,\ \epsilon_g =
0.01$  for individual $b$-jets, $c$-jets and gluon (or light quark)
jets respectively.
We calculate production rates using the MRSD$0'$
parton distributions\cite{mrs} at scale
$Q = m_t$ for both the signal and the backgrounds, assuming $m_t = 150$~GeV
throughout. Since the
$b$-quark distribution is inferred via QCD evolution from
descriptions of deep inelastic scattering data,  there is room for
controversy here; however, both the signal and the ``true" background of
Eq.(\ref{QCDprod}) depend on the same input $b$-distribution. The net signal
and background cross sections, with these cuts and
branching/tagging factors,  are illustrated in Fig.~1 for $pp$ collisions at
$\sqrt s = 14$~TeV.

   Figure~1, which does not include tag-factors,
shows that the charged-Higgs signal has an appreciable size
for some ranges of the parameters $m_{H^\pm}$ and $\tan\beta$.
The $\tan\beta$ dependence is given by a factor $(m_t/\tan\beta)^2
+ (m_b\tan\beta)^2$, with a minimum at $\tan\beta = \sqrt{m_t/m_b}$.
The neighbourhood of this minimum is unpromising for $H^{\pm}$
detection, but many $SUSY$--$GUT$ models suggest that $\tan\beta$
lies near $1$ or alternatively is very large\cite{guts}.
Tagging reduces the major $ttg$ and $ttq$ backgrounds by a factor $1/30$
relative to the signal, making them roughly comparable for
favourable $\tan\beta$.  To improve the signal/background
ratio further and to estimate the mass $m_{H^\pm}$, we propose the
following strategy for event reconstructions.

\renewcommand{\labelenumi}{(\alph{enumi})}
\begin{enumerate}

\item Reconstruct the missing neutrino momentum, by equating
${\bf p}_T(\nu) = \overlay/{\bf p}_T$  and fixing the longitudinal
component $p_L(\nu)$ by the invariant mass constraint  $m(\ell \nu) = M_W$.
The latter gives two solutions in general; if they are complex we discard
the imaginary parts and the solutions coalesce.   We note that the sign
$\pm$ of this $W$ (and hence by inference the other $W$ too) is determined
by the sign of the lepton charge.

\item There are now 6 ways in which two of the $b$-jets can be paired with
the two $W$'s to form top candidates (unless some of the $b$-jets are also
lepton-tagged and thus have known signs).  Together with the two-fold
ambiguity from~(a), this gives 12 candidate reconstructions, in each of
which there are two top mass values $m_{t1},\ m_{t2}$.  We select the
assignment with best fit to the top mass (that will be known),
determined by minimizing $|m_{t1} + m_{t2} - 2m_t|$ subject to the
requirements $|m_{t1} - m_{t2}| < 50$~GeV and $|m_{t1} + m_{t2} - 2m_t|
< 60$~GeV.   If these requirements cannot be met, we reject the event
as unreconstructable.

\item In the selected best-fit assignment above, there are 2 ways in
which the remaining $b$-jet can be paired with one of the top candidates,
so we have 2 candidate values for the reconstructed charged-Higgs mass
$\tilde m_{H \pm} = m(b,t1),\ m(b,t2)$.  Unless the charge of the $b$-jet can
be identified, there is no way
to choose between them  (unless the $b$-jet is also lepton-tagged), so we
retain both values; thus even the signal
events contain an irreducible combinatorial background.  However, the
correct pairings will give a peak in the $\tilde m_{H \pm}$ distribution
while the incorrect pairings and background events will be more broadly
distributed.
\end{enumerate}

This strategy is more ambitious than that of Ref.\cite{gun2}, where
a $b$-jet is combined only with a reconstructed $t \to bjj$ hadronic
system.

Figure~2 compares the signal and background
contributions to the $\tilde m_{H\pm}$ distributions, for
$m_{H\pm} = 200,300,400,500$~GeV  with either $\tan\beta = 1$
or $\tan\beta = 50$; there are two possible values and hence two
counts per event in this graph.  For the most favourable of the
cases illustrated, namely $m_{H\pm} = 200$~GeV with $\tan\beta = 50$,
the signal integrated over the range
 $180 < \tilde m_{H^\pm} < 220$~GeV is 5 counts over a
total background of 4 counts
for each fb$^{-1}$ of luminosity.  With $100\rm\,fb^{-1}$ of luminosity
(one years running at design luminosity $10^{34}\rm\,cm^{-2}\,s^{-1}$)
this signal would be very significant.
As   $ m_{H \pm} $  increases,
both the signal and background fall at comparable rates; for
$m_{H\pm}=500$~GeV, the signal in a 60~GeV bin is $1.0$
over a background of $1.6$~counts/fb$^{-1}$ that would still be
very significant with $100\rm\,fb^{-1}$ luminosity.  If we take
$\tan\beta = 1(2)$ instead,  the background remains essentially the
same while all the signals drop by a factor 2.8(11); hence the regions
$\tan \beta \le 1 $ and $\tan \beta \ge 30 $ are very promising while
the region $2 \le \tan \beta \le 15$ is problematical.  Thus far we
have assumed $m_t = 150$~GeV; for $m_t  = 180$~GeV instead, the
$\tan\beta=1$ signals shown here increase by about $50 \%$ (except
near threshold $m_{H\pm} \sim m_t$) while the net background
falls by about $20 \%$. Lastly we
remark that the assumed cuts above are rather stringent, reducing the
Higgs signal by factors of order 10--30 depending on $m_{H\pm}$,
and the tagging efficiencies may prove to be better than we have
assumed here\cite{dai}; in these respects our event rates may be
viewed as conservative.

    We conclude that the outlook is promising.  With our assumed
tagging efficiencies and cuts, significant $H \to tb$
charged-Higgs signals would be detectable for a limited but
interesting range of the parameters $m_{H\pm}$ and $\tan\beta$.

    We thank Alan Stange and Rahul Sinha for helpful discussions.

\newpage

\section*{Figure captions}

\begin{enumerate}

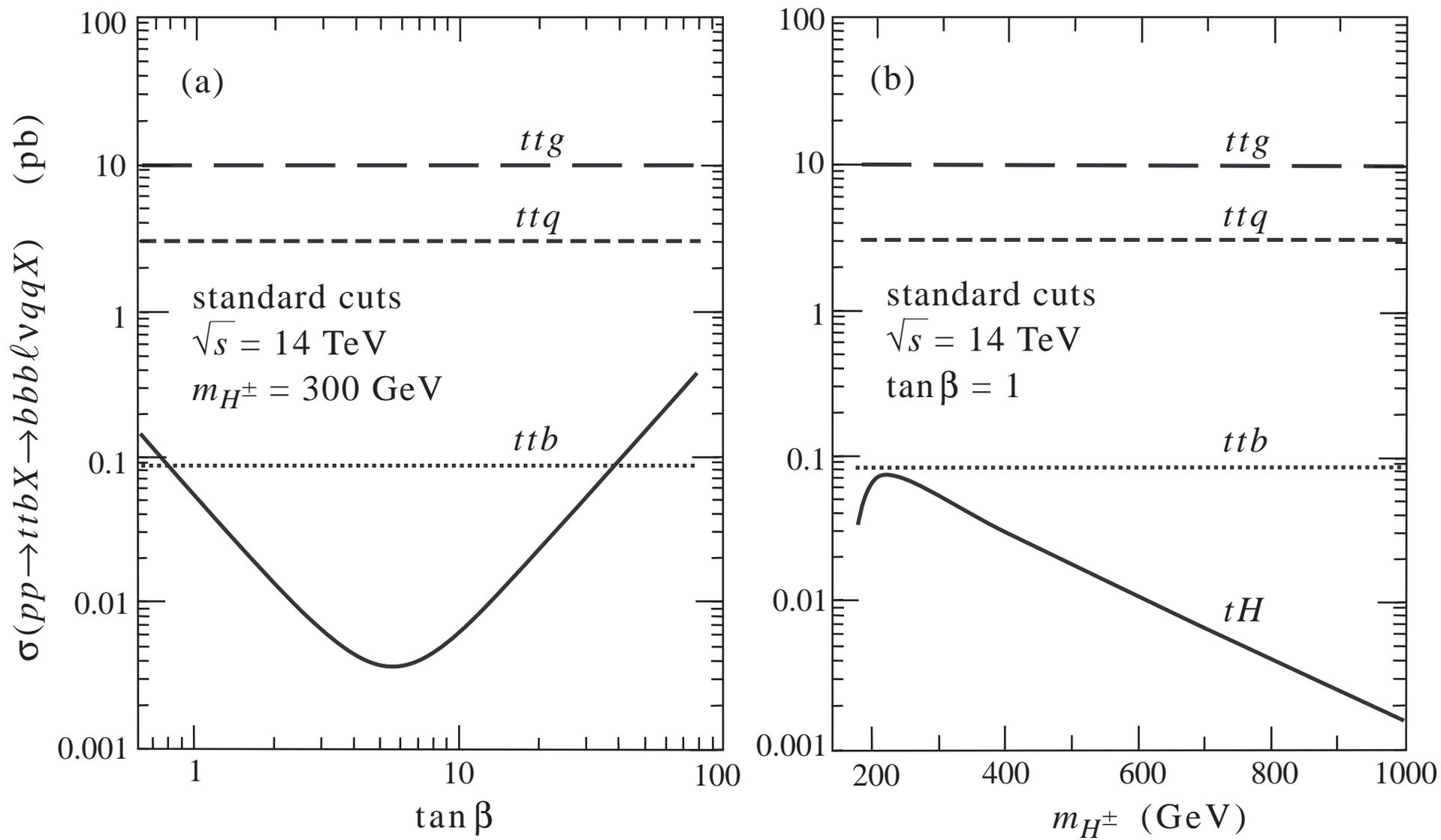
\item[Fig.~1:] Comparison of charged-Higgs signal and principal
backgrounds in the $pp \to ttbX$ channel at $\sqrt s = 14$~TeV,
including branching fractions and acceptance cuts but excluding $b$-tag
factors, with $m_t = 150$~GeV: (a)~cross sections versus $\tan\beta$
for $m_{H\pm} = 300$~GeV; (b)~cross sections versus
$m_{H\pm}$ for $\tan\beta = 1$.

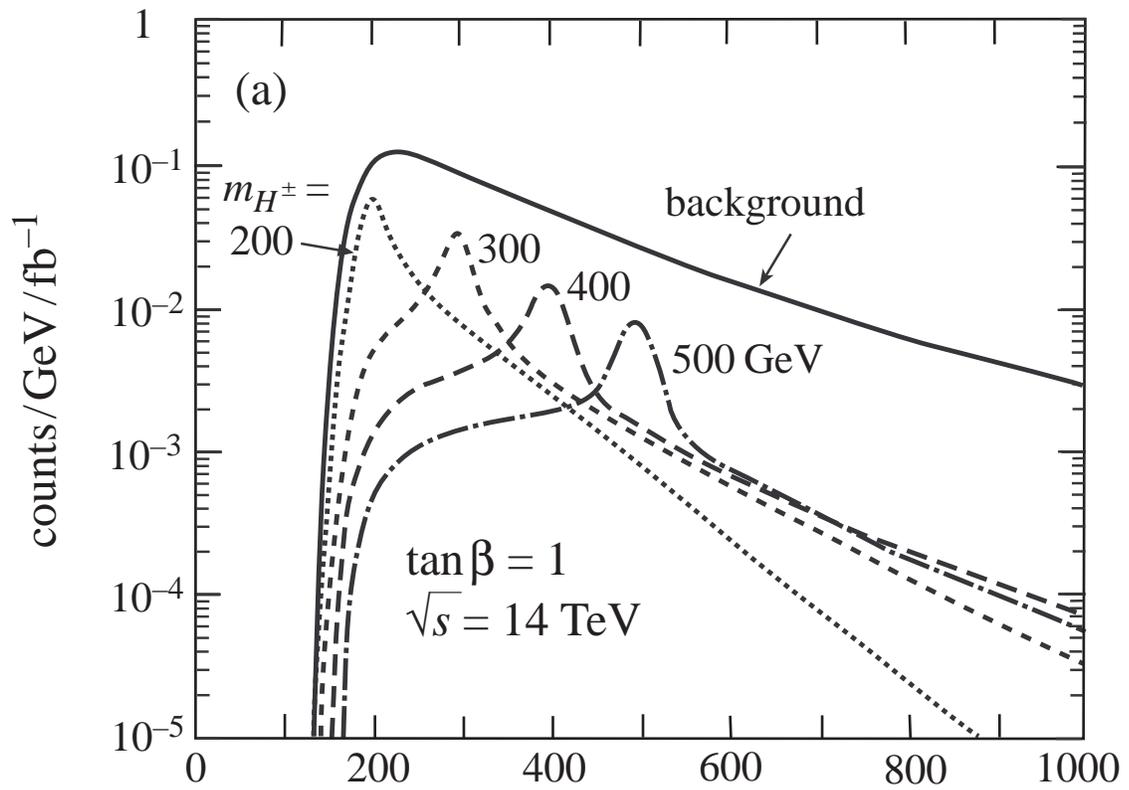
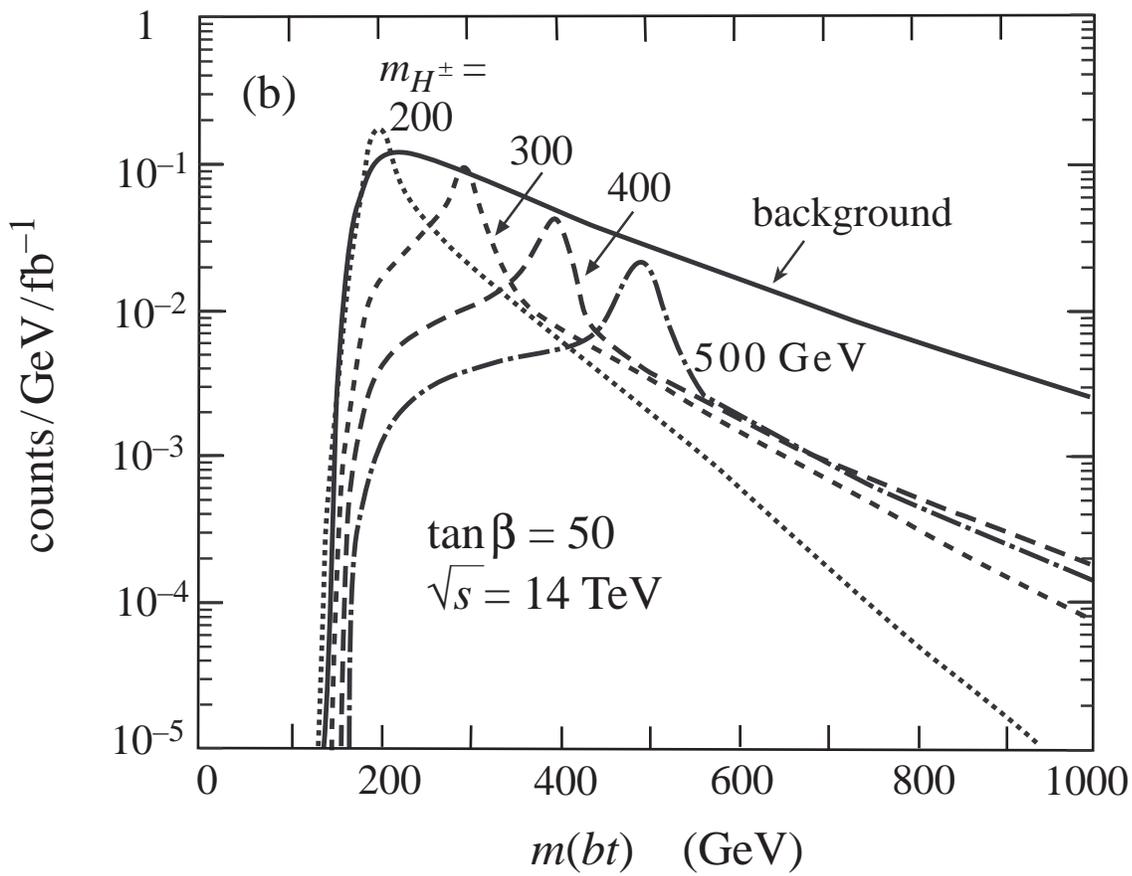
\item[Fig.~2:] Comparison of charged-Higgs signals and summed
backgrounds in the distribution versus  reconstructed charged-Higgs
mass $\tilde m_{H \pm}$, with two counts per event. The cases
$m_{H \pm} = 200,300,400,500$~GeV are shown for (a)~$\tan \beta=1$
and (b)~$\tan \beta = 50$.

\end{enumerate}


\begin{thebibliography}{99}
\frenchspacing

\bibitem{hhg}For a review see J.F.~Gunion, H.~Haber, G.L.~Kane and S.~Dawson,
  ``The Higgs Hunter's Guide" (Addison-Wesley, Reading, MA, 1990).

\bibitem{baer}H.~Baer et al., Phys. Rev. {\bf D46} (1992) 1067.

\bibitem{gunion}J.F.~Gunion, R.~Bork, H.E.~Haber, and A.~Seiden, Phys. Rev.
{\bf D46} (1992) 2040;
J.F.~Gunion, L.H.~Orr, ibid. {\bf D46} (1992) 2052;
J.F.~Gunion, H.E.~Haber, C.~Kao, ibid. {\bf D46} (1993) 2907.


\bibitem{kz}Z.~Kunszt and F.~Zwirner, Nucl. Phys. {\bf B46} (1992) 4914.

\bibitem{barger} V.~Barger, M.S.~Berger, A.L.~Stange, and R.J.N.~Phillips,
Phys. Rev. {\bf D45} (1992) 4128;
V.~Barger, K.~Cheung, R.J.N.~Phillips, and A.L.~Stange, ibid. {\bf D46}
 (1992) 4914.

\bibitem{bhp}V.~Barger, J.L.~Hewett, and R.J.N.~Phillips, Phys. Rev. {\bf D41}
(1990) 3421.

\bibitem{bawa}A.C.~Bawa, C.S.~Kim, and A.D.~Martin, Z.~Phys. {\bf C47}
(1990) 75.

\bibitem{roy}R.M.~Godbole and D.P.~Roy, Phys. Rev. {\bf D43} (1991) 3640;
M.~Drees and  D.P.~Roy, Phys. Lett. {\bf B269} (1991) 155; D.P.~Roy, ibid.
{\bf B277} (1992) 183; ibid. {\bf B283} (1992) 403.

\bibitem{barnett}R.M.~Barnett, J.F.~Gunion, R.Cruz, and B.~Hubbard, Phys. Rev.
{\bf D47} (1993) 1048.

\bibitem{bsg}J.L.~Hewett, Phys. Rev. Lett. {\bf70} (1993) 1045; V.~Barger,
M.S.~Berger, R.J.N.~Phillips, ibid. {\bf 70} (1993) 1368; CLEO collaboration,
report to   Washington APS meeting, April 1993.

\bibitem{bertolini}S.~Bertolini et al., Nucl. Phys. {\bf B353} (1991) 591;
R.~Barbieri   and G.F.~Giudice, Phys. Lett. {\bf B309} (1993) 86.

\bibitem{gkw}J.F.~Gunion, G.L.~Kane, and J.~Wudka, Nucl. Phys. {\bf B299}
 (1988) 231.

\bibitem{dduw} T.~Garavaglia, W.~Kwong, and D.D.~Wu, Phys.\ Rev.\ {\bf D48}
 (1993) 1899.

\bibitem{dai}   J.~Dai, J.F.~Gunion, R.~Vega, Phys. Rev. Lett. {\bf71}
(1993) 2699.

\bibitem{gun2} J.F.~Gunion, work in progress, quoted by J.F.~Gunion
  and S.~Geer, Report of SSC Higgs Working Group, UCD-93-32.

\bibitem{lhc} For latest parameters, see LHC News No.4 and CERN/AC/93-03.

\bibitem{mrs}A.D.~Martin, R.G.~Roberts, and W.J.~Stirling, Phys. Lett.
  {\bf B309} (1993) 492.

\bibitem{guts}e.g. S.~Dimopoulos, L.J.~Hall and   S.~Raby, Phys. Rev. {\bf D45}
 (1992) 4192; V.~Barger, M.S.~Berger, and P.~Ohmann, Phys. Rev. {\bf D47}
 (1993) 1093.

\end{thebibliography}
\end{document}